\documentstyle[preprint,aps,prd]{revtex}
\draft 

\begin{document}

\title{
Collective excitations of a trapped Bose-condensed gas}
 
\author{S. Stringari}
 \address{Dipartimento di Fisica, Universit\`{a} di Trento,}
\address{and Istituto Nazionale di Fisica della Materia,}
\address{I-3850 Povo, Italy}
  
 \date{\today}
 
 \maketitle
 
 \begin{abstract}
By taking the hydrodynamic limit
we derive, at $T=0$, an explicit solution of the linearized 
time dependent Gross-Pitaevskii 
equation for the order parameter of a Bose gas confined
in a harmonic trap and interacting with repulsive forces. 
The dispersion law $\omega=\omega_0(2n^2+2n\ell+3n+\ell)^{1/2}$
for the elementary excitations is obtained, to 
be compared with the prediction 
$\omega=\omega_0(2n+\ell)$ of the noninteracting harmonic oscillator model. 
Here $n$ is the number of radial nodes and $\ell$ is the orbital 
angular momentum.  The effects of the kinetic energy pressure,
neglected in the hydrodynamic approximation, are estimated
using a sum rule approach. Results are also presented 
for deformed traps and attractive forces.
 \end{abstract}
 

\narrowtext
 
 Almost 50 years ago  Bogoliubov \cite{bog} derived his famous theory for the
elementary excitations of a dilute Bose gas. This theory,
originally applied to  homogeneous systems, is now receiving a novel
interest because of the experimental availability of Bose-condensed
gases confined in magnetic traps \cite{boulder,rice,mit} (for a review on
Bose-Einstein condensation see for instance Ref.\cite{levico}). 
The  Bogoliubov theory can be shown \cite{LP} 
to correspond to the linear limit    
of the time dependent Gross-Pitaevskii \cite{GP} equation
for the order parameter $\Phi$
 \begin{eqnarray}
 i\hbar\frac{\partial}{\partial t} \Phi({\bf r},t) =
 \left (-\frac{\hbar^2\nabla^2}{2m} + V_{ext}({\bf r}) +\frac{4\pi \hbar^2 a}{m}
\mid \Phi({\bf r},t)\mid^2 \right )\Phi({\bf r},t) \, .
 \label{GP}
 \end{eqnarray}
Here $V_{ext}$ is the 
confining potential and $a$ is the 
s-wave scattering length. This equation neglects 
interaction effects arising from the atoms out of the 
condensate. This is an accurate approximation for a 
dilute Bose gas at low temperatures where the depletion of the condensate
is negligible. Differently from the homogeneous case, the Gross-Pitaevskii
equation in the presence of an external potential admits stationary solutions
not only for positive values of the scattering length, but also when
$a$ is negative. In the latter case a solution of metastable type
 is found provided the number of atoms in the trap is not too large
\cite{burnett,BP,F,DS}.
The solutions of the  time-dependent Eq.(\ref{GP}), after linearization, have 
the well known RPA structure 
and have been the object of a recent numerical investigation
in the case of a trapped atomic gas \cite{oxford}.

The main purpose of this work is to obtain an explicit, analytic  solution 
of (\ref{GP}) 
holding when the repulsive interaction
is large enough to make the kinetic energy 
pressure negligible compared to the external 
and interparticle interaction terms.
When applied
to the calculation of the ground state this limit
corresponds to the Thomas-Fermi approximation and is reached 
for positive and large values of the adimensional parameter 
$ Na/a_0$ where $a_0 = (\hbar /m\omega_0)^{1/2}$ is the 
harmonic oscillator length characterizing the trap 
and $N$ is the number of atoms.
In the study of the elementary excitations this
approximation corresponds
to the hydrodynamic limit accounting, in homogeneous systems,  for the 
propagation of phonons.

In order to discuss the behavior of the elementary excitations in 
this limit it is convenient to
derive explicit equations for the 
density $\rho({\bf r},t)= |\Phi({\bf r},t)|^2$ and for the 
velocity field
${\bf v}  ({\bf r},t) = \left(\Phi^*({\bf r},t){\bf \nabla} 
\Phi({\bf r},t) - {\bf \nabla}\Phi^*({\bf r},t)\Phi({\bf r},t)\right)/2mi
\rho({\bf r},t)$. These equations can be directly obtained starting from
the time dependent Eq.(\ref{GP})
and take the form
\begin{eqnarray}
 \frac{\partial}{\partial t} \rho + {\bf \nabla} 
({\bf v}\rho) = 0
 \label{continuity}
 \end{eqnarray}
and
\begin{eqnarray}
m \frac{\partial}{\partial t} {\bf v} + 
{\bf \nabla}(\delta\mu + \frac{1}{2}m{\bf v}^2) = 0
\label{Euler}
 \end{eqnarray}
where 
\begin{eqnarray}
\delta \mu = V_{ext} + \frac{4\pi \hbar^2 a}{m}\rho - 
\frac{\hbar^2}{2m \sqrt{\rho}}\nabla^2\sqrt{\rho} - \mu
\label{pressure}
 \end{eqnarray}
is the change of the chemical potential with respect to its ground state
value $\mu$. It is worth noting that these equations do not involve any 
approximation with respect to  
the Gross-Pitaevskii Eq.(\ref{GP}) and hold in the linear as well as
in the non linear regimes. They 
have the general structure  of the dynamic equations 
of superfluids at zero temperature (see for example \cite{NP}). 
In particular Eq.(\ref{Euler})
establishes the irrotational nature  of the superfluid flow.

The density $\rho_0$ relative to the ground state is obtained setting
${\bf v} =0$ and $\delta\mu=0$. This yields the equation
\begin{eqnarray}
V_{ext}({\bf r}) + \frac{4\pi \hbar^2 a}{m}\rho_0 - 
\frac{\hbar^2}{2m \sqrt{\rho_0}}\nabla^2\sqrt{\rho_0} -\mu = 0
\label{gs}
\end{eqnarray}
which, as expected, coincides with the Gross-Pitaevskii equation for the
order parameter $\Phi_0=\sqrt{\rho_0}$ of the ground state. The 
chemical potential $\mu$ is fixed by imposing the proper normalization
to the density $\rho_0$.
In the following we will look for solutions of the above equations
by neglecting the kinetic energy pressure
term $\frac{\hbar^2}{2m \sqrt{\rho_0}}\nabla^2\sqrt{\rho_0}$ in
Eq.(\ref{pressure}).
This approximation yields the well known Thomas-Fermi result
\begin{eqnarray}
\rho_0({\bf r}) = \frac{m}{4\pi \hbar^2 a} 
\left ( \mu - V_{ext}({\bf r}) \right )
\label{tfgs}
\end{eqnarray}
for the ground state density.
The solutions for the time dependent equations are also
easily determined using the same approximation which yields 
the simple expression  $\delta\mu=
4\pi\hbar^2(\rho-\rho_0)/m$ for the change of the chemical potential 
(see Eqs.(\ref{pressure}-\ref{gs})). 
Assuming for simplicity an isotropic harmonic oscillator
potential $V_{ext}({\bf r}) = \omega_0^2r^2/2m$, the equations
of motion (\ref{continuity}-\ref{Euler}), after linearization, 
can be written in the useful form
\begin{eqnarray}
\omega^2\delta\rho = 
-\frac{1}{2}\omega_0^2{\bf \nabla}(R^2-r^2){\bf \nabla}\delta \rho
\label{hds}
\end{eqnarray}
where $\delta
\rho({\bf r})exp(-i\omega t) =
\rho({\bf r},t)-\rho_0({\bf r})$ and $R^2 = 2\mu/m\omega_0^2$ fixes
the boundary of the system where the density 
(\ref{tfgs}) vanishes. In the absence of the external trap
the same procedure yields the well known equation $\omega^2\delta\rho = 
-c^2{\bf \nabla}^2\delta \rho$ where $c=(4\pi \hbar^2 a\rho_0/m^2)^{1/2}$ 
is the sound velocity of the homogeneous Bose gas.

The solutions of the hydrodynamic equations (\ref{hds})
are defined in the interval $0\le r \le R$ and have the form
\begin{eqnarray}
\delta \rho({\bf r}) = {\it P}^{(2n)}_{\ell}(\frac{r}{R})
r^{\ell}Y_{\ell m}(\theta,\phi) 
\label{hdsolution}
\end{eqnarray}
where $P^{(2n)}_{\ell}(t)=1 + \alpha_{2}t^2 + ... + \alpha_{2n}t^{2n}$
are polynomials of degree $2n$, containing
only even powers of $t$, and 
satisfying the orthogonality condition
$\int^1_0 P^{(2n)}_{\ell}({t})P^{(2n^{\prime})}_{\ell}(t)t^{2\ell+2}dt = 0$
if $n\ne n^{\prime}$.
The parameters $\ell$ and $m$ label 
the angular momentum of the excitation and its $z$-component respectively.
The coefficients $\alpha_{2k}$ satisfy the  recurrence relation
$\alpha_{2k+2} = - \alpha_{2k}(n-k)(2\ell+2k+3+2n)/(k+1)(2\ell+2k+3)$.
The dispersion law of the normal modes is given by the formula
\begin{eqnarray}
\omega(n,\ell) = \omega_0(2n^2 + 2n\ell + 3n +\ell)^{1/2}
\label{hdsdispersion}
\end{eqnarray}
which represents the main result of the present work. It
should be compared with the prediction of the harmonic oscillator (HO) model
in the absence of interparticle interactions:
\begin{eqnarray}
\omega_{HO} = \omega_0(2n + \ell) \, \, .
\label{hodispersion}
\end{eqnarray}

Of particular interest is the case of  the lowest radial modes ($n=0$),
also called surface excitations, for which we predict
the dispersion law
\begin{eqnarray}
\omega(n=0) = \sqrt{\ell}\omega_0 \, .
\label{n=0dispersion}
\end{eqnarray}
The frequency of these modes lies systematically below the 
harmonic oscillator result
$\omega_{HO}(n=0) = \ell \omega_0$. This behavior should be taken into 
account in the determination of the critical frequency
$\omega_{cr}=min_{\ell}(\omega(\ell)/\ell)$ needed to 
generate a rotational instability
\cite{DS,SS}, and might provide a competitive mechanism
with respect to the creation of a
vortex line.  Notice that in the dipole case ($\ell=1$)
both the hydrodynamic and harmonic oscillator 
predictions coincide with the oscillator frequency $\omega_0$. 
This follows
from the fact that in an external harmonic potential the lowest dipole mode
corresponds to a rigid motion of the center of mass, and is consequently
unaffected by the interatomic forces.

The accuracy of prediction (\ref{hdsdispersion}) 
is expected to become lower and lower as $n$ and $\ell$ 
increase. In fact the
high energy states are associated with rapid variations of the density in space
and consequently the kinetic energy 
contribution in Eq.(\ref{pressure}) cannot be longer neglected. Using the 
macroscopic  language
this corresponds to abandon the phonon
regime. The energy range where our prediction (\ref{hdsdispersion}) 
is expected to be accurate 
then corresponds to values smaller than the chemical 
potential $\mu$.
For large values of $n$ and $\ell$, the correct dispersion law will 
approach the harmonic oscillator  result (\ref{hodispersion}).

The effects of the kinetic energy pressure, ignored in the hydrodynamic
approximation,
can be investigated by calculating the 
energy of the collective mode through a sum rule approach \cite{sumrules}, 
based on the ratio
\begin{eqnarray}
\hbar^2\omega^2 = \frac{m_3}{m_1} 
\label{omega31}
\end{eqnarray}
between the cubic energy weighted and the energy weighted
moments of the  dynamic structure factor: 
$m_p=\sum_n\mid<0\mid F\mid n>\mid^2\hbar\omega_{n0}^p$. Here $\hbar\omega_{n0}$
is the excitation energy of the state $|n>$ 
and $F$ is a general excitation operator. 
The moments $m_1$ and $m_3$ can be reduced in the form of commutators involving
the hamiltonian of the system. One finds
$m_1={1\over 2}<0[F^{\dagger}[H,F]]0>$
and $m_3 = {1\over 2}<0[[F^{\dagger},H],[H,[H,F]]]0>$. 
Equation (\ref{omega31}) provides in general  
a rigorous upper bound to the energy of
the lowest state excited by $F$. 
The sum rule approach has the merit of providing useful information
on the dynamic behavior of the system using only the knowledge of the ground 
state.
By evaluating explicitly the commutators 
with the Hamiltonian \cite{effective}
\begin{eqnarray}
H = \sum_i{p^2_i\over 2m} + \sum_i{1\over 2}m\omega_0^2r^2_i +
{4\pi\hbar^2 a\over m}\sum_{i<j}\delta({\bf r}_i-{\bf r}_j) 
\label{H}
\end{eqnarray} 
we find, in the case of the surface ($n=0$)
operator $F=\sum_ir_i^{\ell}Y_{\ell m}(\theta_i,\phi_i)$, 
the results
\begin{eqnarray}
m_1 = \frac{\hbar^2}{8m\pi}\ell(2\ell+1)\int r^{2\ell-2}\rho d{\bf r}
\label{m1}
\end{eqnarray}
and
\begin{eqnarray}
m_3 = \frac{\hbar^4}{8m^2\pi}\ell(2\ell+1)
\left[\ell m\omega_0^2\int r^{2\ell-2}\rho d{\bf r}
+\ell(\ell-1){\hbar^2\over m}\int\mid{\bf \nabla}\sqrt{\rho}\mid^2r^{2\ell-4} 
d{\bf r}\right] \, \, ,
\label{m3}
\end{eqnarray}
and the excitation frequency (\ref{omega31}) takes the form
\begin{eqnarray}
\omega^2(n=0) = \omega_0^2\ell\left(1+(\ell-1)\beta_{\ell}\right)
\label{omegabound}
\end{eqnarray}
where 
$\beta_{\ell}= \hbar^2\int\mid{\bf \nabla}\sqrt{\rho}\mid^2r^{2\ell-4}d{\bf r}/
m^2\omega_0^2\int r^{2\ell-2}\rho d{\bf r}$.
For the most relevant quadrupole ($n=0,\ell=2$) case we find 
\begin{eqnarray}
\omega_Q = \sqrt2\omega_0(1 + \frac{E_{kin}}{E_{ho}})^{1/2}
\label{omegaqua}
\end{eqnarray}
where $E_{kin}$ and $E_{ho}$ are, respectively, the expectation value of
the kinetic and harmonic potential energies in the ground state. 
In the absence of interparticle interactions one has 
$E_{kin}=E_{ho}$ and hence one recovers
the  harmonic oscillator  result $\omega=2\omega_0$ of 
Eq.(\ref{hodispersion}).
Viceversa when the interaction is repulsive and the number of atoms
is sufficiently large, the kinetic energy term is negligible (Thomas-Fermi
limit) and one obtains the hydrodynamic 
prediction $\omega=\sqrt2\omega_0$ of eq.(\ref{n=0dispersion}).
The knowledge of the  kinetic energy relative to the ground state then permits
to estimate the quadrupole excitation energy in the general case.
For higher multipolarities the determination of the coefficient $\beta_{\ell}$
of Eq.(\ref{omegabound}) requires the knowledge of finer details
of the ground state. A simple estimate can be obtained 
using the gaussian approximation for
the  wave function of the condensate, yielding $\beta_{\ell}=
E_{kin}/E_{ho}$, independent of $\ell$.

In a similar way, starting from eq.(\ref{omega31}), 
one can determine the frequency
of the compression modes. For the lowest monopole mode
$(n=1,\ell=0)$, excited by the 
operator $F=\sum_ir^2_i$, we find 
\begin{eqnarray}
m_1 = \frac{2\hbar^2}{m}N<r^2>
\label{m1M}
\end{eqnarray}
and
\begin{eqnarray}
m_3 = \frac{2\hbar^4}{m^2}(4E_{kin}+4E_{ho}+9E_{int})
\label{m3M}
\end{eqnarray}
where $E_{int} = {2\pi \hbar^2 a\over m}\int d{\bf r} \rho^2_0(r)$ 
is the interaction energy.
Using the virial identity $2E_{kin} - 2E_{ho} + 3E_{int} = 0$, holding
for the ground state, we finally obtain the following
result for the monopole frequency
\begin{eqnarray}
\omega_M = \omega_0(5-\frac{E_{kin}}{E_{ho}})^{1/2}
\label{omegamon}
\end{eqnarray}
yielding the values $\omega=2\omega_0$ and $\omega=\sqrt5\omega_0$
in the harmonic oscillator and Thomas-Fermi limits respectively. 

Predictions (\ref{omegaqua},\ref{omegamon})
well agree with the results 
recently obtained in Ref.\cite{oxford} 
by solving numerically the time dependent 
Gross-Pitaevskii equations for a trapped atomic gas interacting
with repulsive forces. The conditions of Ref.\cite{oxford} 
correspond to  the value $\mu=4.3\omega_0$ for the chemical potential and
to the value $0.18$ for the ratio between the kinetic and harmonic 
oscillator energies.
This yields, using eqs.(\ref{omegaqua}) and (\ref{omegamon}), the values
$\omega_Q=1.54\omega_0$ and $\omega_M=2.20\omega_0$ in excellent
agreement with the findings of \cite{oxford} 
($1.53\omega_0$ and $2.19\omega_0$ respectively).

The above results for the dispersion law can be 
generalized to the case of a deformed trap.
This is particular relevant since the available 
magnetic traps are often highly anisotropic.
Let us consider the case of a harmonic oscillator
trap with axial symmetry along the z-axis: $V_{ext} = 
m\omega_{\perp}^2s^2/2 + m\omega_z^2z^2/2$ where
$s=(x^2+y^2)^{1/2}$ is  the radial variable in the $x-y$ plane.
In this case the relevant differential equation (\ref{hds}) should be replaced
by 
\begin{eqnarray}
\omega^2\delta\rho = -{1\over 2}
{\bf \nabla}\left(\omega_{\perp}^2(S^2-s^2)
+\omega_z^2(Z^2-z^2)\right){\bf \nabla}\delta \rho
\label{hddeformed}
\end{eqnarray}
with $m \omega_{\perp}^2S^2/2 = 
m\omega_z^2Z^2/2 \equiv \mu$.

Due to the axial symmetry of the trap the third component $m$ of the angular
momentum is  still a good quantum number. However the dispersion law 
will depend on $m$.
Explicit results are available in some
particular cases. For example functions of the form 
$\delta\rho = r^\ell Y_{\ell m}(\theta,\phi)$ are still solutions
of eq.(\ref{hddeformed}) for $m=\pm \ell$ and $m=\pm(\ell-1)$. The resulting
dispersion laws are:
\begin{eqnarray}
\omega^2(m=\pm \ell) = \ell\omega_{\perp}^2
\label{pml}
\end{eqnarray}
and
\begin{eqnarray}
\omega^2(m=\pm (\ell-1)) = (\ell-1)\omega_{\perp}^2 +\omega_z^2 \, .
\label{pml-1}
\end{eqnarray}

Equations (\ref{pml}-\ref{pml-1}) provide a full description of 
the dipole excitation ($\ell=1$)
whose frequencies coincide, as expected, with the unperturbed harmonic 
oscillator values $\omega_D(m=\pm 1)=\omega_{\perp}$ and 
$\omega_D(m=0)=\omega_z$. Viceversa for the quadrupole ($\ell=2$) mode 
Eqs.(\ref{pml}-\ref{pml-1}) account only for the 
$m=\pm 2$ ($\omega=\sqrt2\omega_{\perp}$) and $m=\pm 1$ 
($\omega=(\omega_{\perp}^2+\omega_z^2)^{1/2}$) components.
The solution with $m=0$ involves a coupling with the monopole ($n=1,\ell=0$) 
excitation and 
the dispersion law of the two decoupled modes is given by
\begin{eqnarray}
\omega^2(m=0) = \omega_{\perp}^2\left( 2 + \frac{3}{2}\lambda^2 \mp\frac{1}{2} 
\sqrt{9\lambda^4-16\lambda^2+16}\right)
\label{pml-2}
\end{eqnarray}
with $\lambda = \omega_z/\omega_{\perp}$. When $\lambda \to 1$
one recovers the original solutions  
(\ref{omegaqua}) and (\ref{omegamon}) corresponding, respectively, to the 
quadrupole and monopole excitations in a spherical trap.

For systems interacting with attractive forces ($a<0$) the hydrodynamic results
discussed above do not provide an adequate description of the 
dispersion law and in this case the sum rule approach becomes 
particularly useful.
The kinetic energy contribution $E_{kin}$ entering 
Eqs.(\ref{omegaqua},\ref{omegamon}) can never be neglected,
being larger than $E_{ho}$, and the monopole mode turns out to be located
below the quadrupole 
one. Physically this reflects the tendency of the system to become
more compressible. 
In Fig.2 we show the behavior of the monopole and quadrupole frequencies 
obtained 
using Eqs.(\ref{omegaqua},\ref{omegamon}) as a function of the adimensional
parameter $Na/a_0$. The ratio 
$E_{kin}/E_{ho}$ has been estimated using a variational calculation
of the ground state based on gaussian trial wave functions for
the order parameter \cite{BP,F}.
One can see that while
the energy of the quadrupole excitation 
is enhanced at negative scattering lengths,
the monopole mode becomes softer. 
The upper bound (\ref{omegamon}) for the monopole energy
vanishes when $E_{kin}=5E_{ho}$. This exactly coincides with the condition
for the onset of instability predicted by the use of gaussian 
trial wave functions in the variational calculation ($Na/a_0=-0.67$ \cite{F}). 
It is however worth noting that
the results shown in the figure provide only a semi-quantitative
estimate of the excitation frequencies. In fact
the onset of instability obtained from the exact
solution \cite{burnett} of the Gross-Pitaevskii equation ($Na/a_0 \sim - 0.57$)
differs from the gaussian estimate. Furthermore one should keep in
mind that the estimates ($\ref{omegaqua},\ref{omegamon}$),
being based on the ratio (\ref{omega31}), provide only an
upper bound to the frequency of the lowest excitations.

In conclusion we have derived a systematic investigation
of the collective excitations of a Bose condensed gase 
confined in an external trap.
We have obtained analytic results for the dispersion law of
both surface and compression modes employing the hydrodynamic 
approximation and the sum approach.
 Our work reveals the key role played 
the interatomic forces which introduce a rich structure
in the dynamic behaviour of these new many-body systems.

Stimulating discussions with Lev Pitaevskii are acknowledged. I am also
indebted to Keith Burnett, Mark Edwards and Alexander Fetter for 
useful information on their works.

\begin{figure}
\caption{Quadrupole (Q) and monopole (M) 
frequencies as a function of the adimensional 
parameter $Na/a_0$, calculated using Eqs.(17) and (20).}

\end{figure}
  

\begin{references}
\bibitem{effective} The interaction term in Eq.(\ref{H}) has the meaning
of an effective potential to be employed in the framework of the mean field 
scheme yielding the Gross-Pitaevskii Eq.(\ref{GP}).
\end{references}
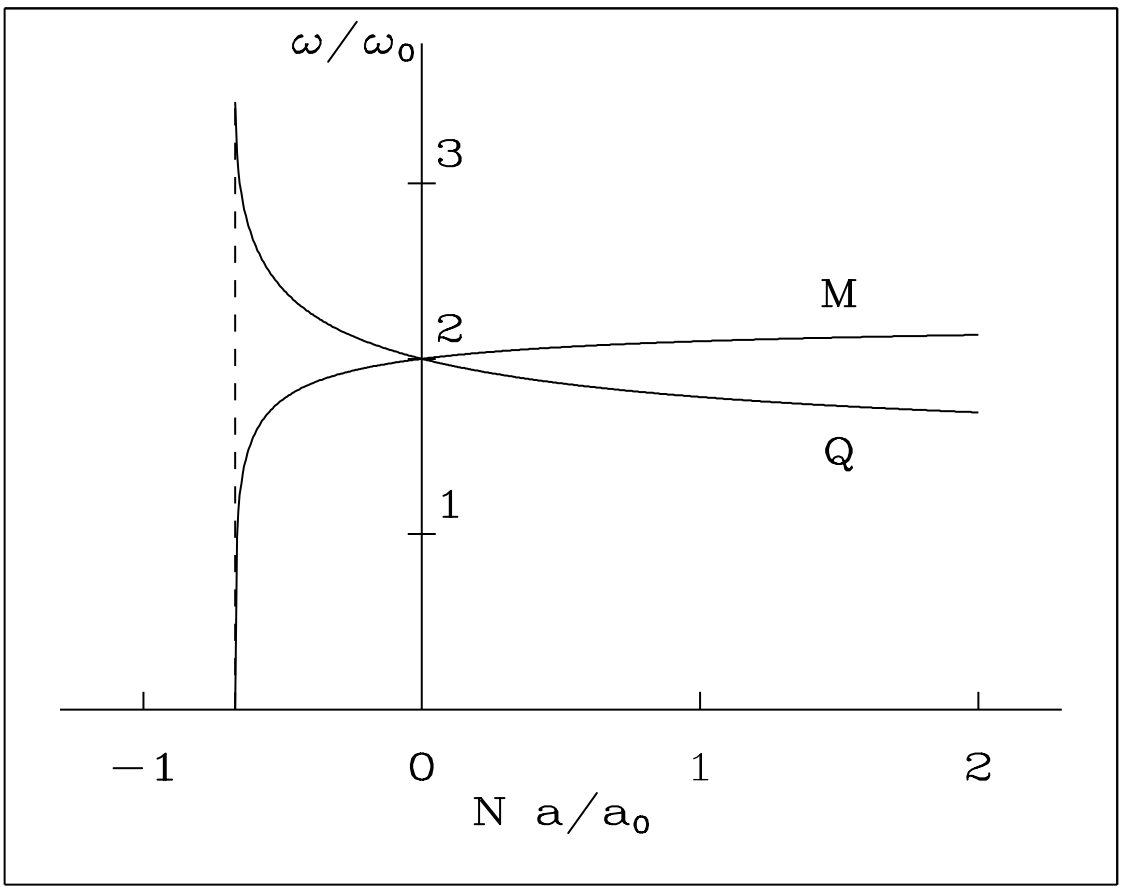
\end{document}